\RequirePackage{fix-cm}

\documentclass[%
twocolumn, %
nofootinbib,
]{revtex4-1}
\usepackage{cmap}

\usepackage{ucs}
\usepackage[utf8x]{inputenc}
\usepackage[T1,T2A]{fontenc}
\usepackage[german,russian,english]{babel}

\usepackage[sort&compress]{natbib}
\usepackage{amsmath}
\usepackage{amssymb}

\usepackage{mathtools}
\mathtoolsset{
showonlyrefs,
mathic = true
}

\allowdisplaybreaks

\usepackage{hyperref}
\hypersetup{backref,
 colorlinks=false}
\hypersetup{pdfborder=0 0 0}

\usepackage{breakurl}

\usepackage{microtype}
\UseMicrotypeSet[protrusion]{alltext}

\usepackage{fixltx2e}

\usepackage{nicefrac}

\usepackage{physymb}

\makeatletter
\def\ps@pprintTitle{%
     \let\@oddhead\@empty
     \let\@evenhead\@empty
     \let\@oddfoot\@empty
     \let\@evenfoot\@oddfoot}
\makeatother
\usepackage{youngtab}

\begin{document}
\title{Tensor computations in computer algebra systems}

\author{A. V. Korolkova}
\email{akorolkova@sci.pfu.edu.ru}
\author{D. S. Kulyabov}
\email{yamadharma@gmail.com}
\author{L. A. Sevastyanov}
\email{leonid.sevast@gmail.com}

\affiliation{Telecommunication System Department\\
Peoples' Friendship University of Russia\\
Miklukho-Maklaya str., 6, Moscow, 117198, Russia}

\thanks{Published in:
  A.~V. Korol’kova, D.~S. Kulyabov, and L.~A. Sevast’yanov.
 {Tensor computations in computer algebra systems}.
 \emph{Programming and Computer Software}, 39\penalty0 (3):\penalty0
 135--142, 2013.
 ISSN 0361-7688.
 \href{http://dx.doi.org/10.1134/S0361768813030031}{doi:~10.1134/S0361768813030031}.}

\thanks{Sources:
  \url{https://bitbucket.org/yamadharma/articles-2011-cas_tensor}}

\begin{abstract}
This paper considers three types of tensor computations. On their
basis, we attempt to formulate criteria that must be satisfied by a
computer algebra system dealing with tensors. We briefly overview the
current state of tensor computations in different computer algebra
systems. The tensor computations are illustrated with appropriate
examples implemented in specific systems: Cadabra and Maxima.
\end{abstract}

\maketitle

\section{Introduction}

Tensor calculations are used in many fields of physics. It should be
noted that the formalism of tensor analysis manifests itself in all
its might not in all areas; rather frequently, its simplified versions
are used.

Each tensor operation itself is sufficiently simple. However, even
standard computations have to involve many elementary
operations. These operations require great care and thoroughness. That
is why it is important in this field to use different simplified
notations and optimized operations (for example, Penrose tensor
diagrams).

One of the tasks of computer algebra systems is to free the researcher
from routine operations, which is also important in the case of tensor
calculus.

\section{Main Application Fields and Types of Tensor Notations}

To define the key operations with tensors, consider the main field of
application.

\subsection{Nonindex Computations for Theoretical Constructs}

Nonindex computations are commonly used in theoretical constructs and
often opposed to component computations. We implement the main tensor
operations in the nonindex case.

\begin{itemize}

\item \emph{Addition of tensors}. The addition of two tensors of
  valence $\left[\begin{smallmatrix} p\\q \end{smallmatrix}\right]$
  yields a tensor of valence $\left[\begin{smallmatrix}
    p\\q \end{smallmatrix}\right]$:
\begin{equation}
  \label{eq:9}
  A + B = C.
\end{equation}
The addition of tensors gives the structure of an Abelian group.

\item \emph{Multiplication of tensors}. The multiplication of tensor $A$ with valence $\left[\begin{smallmatrix}
      p\\q \end{smallmatrix}\right]$ by tensor $D$ with valence
  $\left[\begin{smallmatrix} r\\s \end{smallmatrix}\right]$ yields
  tensor $E$ with valence $\left[\begin{smallmatrix}
      p+r\\q+s \end{smallmatrix}\right]$:
\begin{equation}
  \label{eq:11}
  A \otimes D = E.
\end{equation}
The tensor multiplication defines the structure of a noncommutative
semigroup.

\item \emph{Operation of Contraction}. Let us denote the operation of
  contraction of tensors with respect to last indices by
  $\mathfrak{C}$. Then, under the action of this operation, tensor $F$
  with valence $\left[\begin{smallmatrix}
      p+1\\q+1 \end{smallmatrix}\right]$ goes into tensor $G$ with
  valence $\left[\begin{smallmatrix} p\\q \end{smallmatrix}\right]$:
\begin{equation}
  \label{eq:12}
  \mathfrak{C} F = G.
\end{equation}

\item \emph{Operation of Permutation of Indices}. This operation is necessary for
specifying the symmetry of tensors (for example, a tensor commutator
or anticommutator), for expanding the operation of contraction on the
contraction with respect to arbitrary indices. However, the nonindex
approach gives no way of identifying this operation. Yet, the simplest
symmetries can be explicitly specified in the object description (in
this case, one has to impose restrictions on valence to ensure
uniqueness).

\end{itemize}

\subsection{Vector Computations}

Vector calculus is the simplest case of tensor calculus (a vector is a
tensor of valence one). The $N$-dimensional vector $a^{n}$ is represented
as a set of components $n=\overline{1,N}$, depending on the basis, and a
linear law of the transformation of components for a changing
basis. The frequently used operations are the construction of various
differential operators and change of the basis. The most common
operators are gradient, divergence, and curl (specific to the
three-dimensional space $\mathbb{R}^3$)~\cite{pfur-2012-1,pfur-2011-2::en}).

Component calculations require a basis, a metric, and connectivity
(and, accordingly, a covariant derivative) to be defined. In vector
calculus, it is common to use a holonomic basis, which is constructed
as a set of partial derivatives of the coordinates in a tangent bundle
and the dual basis as a 1-form in the cotangent bundle:
\begin{equation}
  \label{eq:2}
  \vec{\delta}_{i} = \frac{\partial}{\partial x^{i}},
  \quad
  \vec{\delta}^{i} = d x^{i}.
\end{equation}
The connectivity and metric are constructed in such a way that the
covariant derivative of the metric be equal to zero:
\begin{equation}
  \label{eq:3}
  \nabla_{k} g_{i j} = 0.
\end{equation}
In this case, the connectivity and metric are consistent~\cite{mangiarotti2000connections}.

It should be noted that vector computations often use a special
nonholonomic basis that makes it possible not to distinguish between
contravariant and covariant vectors and keep the dimension unchanged
under a coordinate transformation\footnote{Under this
transformation, length goes into length, angle goes into angle, etc.}
(for details, see~\cite{pfur-2012-1}):
\begin{equation}
  \label{eq:4}
  \vec{\delta}_{i} = \frac{\partial}{\partial s^{i}},
  \quad
  \vec{\delta}^{i} = d s^{i},
  \quad
  d s^{i} = h_{j}^{i} d x^{j}.
\end{equation}
Here, $d s^{i}$ is the element of length with respect to a given
coordinate and $h_{j}^{i}$ are the nonholonomy coefficients (the Lam\'{e}
coefficients in the case of orthogonal coordinates).

\subsection{Dirac 4-Spinors}

A special case of tensor objects are spinors (also called
spin-tensors). Particularly, spinors are representations of the
Lorentz group with a half-integer highest weight. Conventional tensors
are representations with an integer highest weight.

For historical reasons, the investigations most commonly use Dirac
4-spinors, which are applied to write the Dirac equations describing
fermions with spin ~$\frac{1}{2}$. Dirac 4-spinors are essentially irreducible
spinors for the case $n = 4$ and $s=\pm 2$, where $n$ is the dimension of
the vector space, $s = n - 2u$ is its signature, and $u$ is the number
of negative values of the diagonal metric tensor $g_{a b}$.

Usually the handling with Dirac spinors is based on $\gamma$-matrices
derived from the Clifford--Dirac equation~\cite{cartan}:
\begin{equation}
  \label{eq:5}
  \gamma_{(a} \gamma_{b)} = g_{a b} \hat{I},
\end{equation}
where $\gamma_{a}$ are $N \times N$ matrices, $g_{a b}$ is a metric
tensor, $\hat{I}$ is
the identity $N \times N$ matrix, and $N$ is the dimension of
the spinor space:
\begin{equation}
  \label{eq:7}
  N = 
  \begin{cases}
    2^{n/2}, &\text{even $n$,} \\
    2^{n/2 - 1/2}, &\text{odd $n$.}
  \end{cases}
\end{equation}
where $\gamma$-matrices are Clifford algebra elements generating a
linear transformation of the spin space.

Since the $\gamma$-matrix can be regarded as the coefficients of
transition from spin to vector space, the spin coefficients should be
introduced more strictly and Eq.~\eqref{eq:5} should be written in the
following way
\begin{equation}
  \label{eq:6}
  \gamma_{a \rho}^{\sigma} \gamma_{b \sigma}^{\tau} + \gamma_{b
    \rho}^{\sigma} \gamma_{a \sigma}^{\tau} = 2 g_{a b} \delta_{\rho}^{\sigma}.
\end{equation}

The construction of a complete Clifford algebra requires also products
of $\gamma$-matrices, but in view of~\eqref{eq:5}, it is suffice to consider
only antisymmetrized products
\begin{equation}
  \label{eq:10}
  \gamma_{a b \dots d} := \gamma_{[a} \gamma_{b} \cdots \gamma_{d]}. 
\end{equation}
Also, we introduce an element $\gamma_{5}$:
\begin{equation}
  \label{eq:8}
  \gamma_{5} := \frac{i}{4!} e^{a b c d} \gamma_{a}\gamma_{b}\gamma_{c}\gamma_{d},
\end{equation}
where $e^{a b c d}$ is an alternating tensor.

The handling with $\gamma$-matrices is reduced to a set of relations
following from algebraic symmetries, such as
\begin{gather}
    \gamma^a \gamma_a = 4 \hat{I}, 
    \label{eq:gamma:sym:1} \\
    \gamma^a \gamma^b \gamma^c \gamma_a = 4 g^{b c} \hat{I} ,
    \label{eq:gamma:sym:2} \\
    \gamma_{a} \gamma_{b} = \gamma_{ab} + g_{ab} \hat{I},
    \label{eq:gamma:sym:3} \\
    \gamma_a \gamma_b \gamma_c = \gamma_{a b c} +
    g_{a b} \gamma_c + g_{b c}\gamma_a -
    g_{a c} \gamma_b.   
    \label{eq:gamma:sym:4}
\end{gather}

\subsection{Tensor Calculations in the Theory of General Relativity}

General relativity became the first physical theory requiring the
whole power of differential geometry and tensor
calculations~\cite{gerdt:1980:ufn::en}. These calculations involve bulky
tensor constructs that can be
simplified in view of the symmetry of tensors. Usually, one
differentiates between monoterm and multiterm symmetries. A key
element of the theory is the Riemann tensor, which has both very
simple monoterm and complex multiterm symmetries like the Bianchi
identities.

Monoterm symmetries correspond to simple permutation symmetries and
are given by a group of permutations. Specifically, for the Riemann
tensor, we have
\begin{equation}
  \label{eq:riman:1}
  R_{bacd} = - R_{abcd}, \quad R_{cdab} = R_{abcd}.
\end{equation}

Multiterm symmetries are given by an algebra of permutations. The
Bianchi identity has the form\footnote{The round brackets
  in~\eqref{eq:riman:2} denote symmetrization.}:
\begin{equation}
  \label{eq:riman:2}
R_{a(bcd)} = R_{abcd} + R_{acdb} + R_{adbc} = 0.
\end{equation}
The differential (second) Bianchi identity has the
form\footnote{Semicolon in~\eqref{eq:riman:3} denotes covariant
  derivative.}:
\begin{equation}
  \label{eq:riman:3}
 R_{ab(cd;e)} = \nabla_{e} R_{abcd} + \nabla_{c} R_{abde} + \nabla_{d} R_{abec} = 0.
\end{equation}

It is most reasonable to specify symmetries by means of Young
diagrams~\cite{barut:1986::en}. Here, the presence of predefined classes
of tensors does not
eliminate the need for an explicit specification of symmetry. For
example, the Riemann tensor $R_{abcd}$ in different sources has the
symmetries {\scriptsize\young(ac,bd)} and {\scriptsize\young(ab,cd)}.

\subsection{Types of Tensor Notation}

Thus, based on the above types of tensor calculations, one can specify
three types of tensor notation: component notation, notation with
abstract indices, and nonindex notation. Each type has its own
specificity and application field.

Component indices actually turn a tensor into a set of scalar values
used in specific calculations. Usually, it makes sense to operate with
component indices only after simplifying the tensor expression and
taking into account all of its symmetries.

The nonindex notation is often used when the researcher is interested
in the symmetry of tensors rather than in the final result. However,
this form of notation lacks in its expressiveness: the tensor is
regarded as an integral entity; accordingly, only the symmetries that
are related to the tensor as a whole can be considered. To operate
with objects of complex structures, one has to invent new notations or
add verbal explanations. It is this problem that should be treated by
abstract indices~\cite{penrose::en}.

Abstract indices should be regarded as an improvement of the nonindex
notation of tensors. An abstract index denotes merely the fact that a
tensor belongs to a certain space, rather than obeying the tensor
transformation rule (unlike component indices). In this case, one can
consider both symmetries covering the full tensor (all its indices)
and symmetries of individual groups of indices.

\section{Tensor Computations and Computer Algebra Systems}

Modern computer algebra systems are able to solve problems of a
sufficiently wide class and from different areas of knowledge. The
systems can be highly specialized or with a claim to universality (a
survey of some systems can be found, for example,
in~\cite{cain,pfur-2007-1-2::en,ech::en}). We
consider some computer algebra systems that to some extent can operate
with tensors.

\subsection{Requirements for Computer Algebra System}

Three types of tensor notations correspond to three types of tensor
analytical calculations; this leads to certain requirements for
computer algebra systems.

Nonindex computing handles with tensors as integral algebraic
objects. In this case, one can either specify the simplest type of
symmetry (the object is a representation of a group or algebra) or use
objects with predefined symmetry.

Abstract indices require the ability of specifying complex types of
symmetry, for example, through Young diagrams. In addition, it is
necessary to be able to work with dummy indices, specify and consider
them when bringing into canonical form. Both types of abstract
computations use information about symmetries for bringing into
canonical form and simplifying tensor expressions.

Component indices require in fact a scalar system of computer algebra
and possibly the presence of simple matrix operations. In fact, a
specific coordinate system and metric are given. Since all operations
are performed by components, the information about the tensor as an
integral object and about its symmetries is lost. Therefore, all
operations with symmetries and bringing into canonical form must be
carried out in the previous phase of the study.

\subsection{Notation}

The use of computer algebra systems often implies interactive
functioning of users. In this case, the convenience of notation plays
a key role. Historically, the mathematical notation of tensors follows
the notation of the \TeX{} system; namely, the tensor $T^{a}_{b}$ is written
as \verb|T^{a}_{b}|. Therefore, the use of this notation would be quite
natural. This approach was implemented in \emph{Cadabra} however, this is
a specialized system for tensor computing. A tensor notation for
general-purpose computer algebra systems should account for the
limitations of these systems (for example, the symbol \verb|^| is normally
reserved and used for exponentiation).

Because general-purpose systems operate with functions and the basic
internal data structure is a list, a functional--list notation is
used. The name of a function can be given by the tensor name, and the
covariant and contravariant indices are given either by a prefix (for
example, as in \emph{xAct}):
\begin{verbatim}
T(a,-b),
\end{verbatim}
or positionally (for example, as in the \emph{Maxima}):
\begin{verbatim}
T([a],[b]).
\end{verbatim}
It is also possible to use associative lists, such as
\begin{verbatim}
Tensor[Name["T"], Indices[Up[a], Down[b]]].
\end{verbatim}

Let us consider the most interesting (for practical use)
implementations of tensor calculations in different computer algebra
systems.

\subsection{Cadabra}

Cadabra [\url{http://cadabra.phi-sci.com/}] refers to the type of
specialized computer algebra systems. The area of its specialization
is the field theory. Because complex tensor calculations are an
integral part of the field theory, it is not surprising that tensor
calculations in this system are supported at a high level.

However, the field theory operates mostly with abstract indices, and
component computations receive much less attention. Most likely, this
is the reason that component computations have not yet been
implemented in Cadabra, although this option has been projected for
implementation.

However, component computations require that a computer algebra system
possess the capabilities of general-purpose systems, which cannot be
found in Cadabra.

\subsection{Maxima}

Maxima [\url{http://maxima.sourceforge.net/}] is one of the major freeware
general-purpose computer algebra systems. Maxima was derived from
Macsyma, a system developed in MIT from 1968 to 1982.

Maxima implements all the three types of tensor calculations~\cite{toth}:
\begin{itemize}
\item package \emph{atensor}---nonindex algebraic calculations (with a set of
basic algebras and main purpose of simplifying tensor expressions by
manipulations with both monoterm and multiterm symmetries);

\item package \emph{ctensor}---component calculations (with an option of
manipulating with metrics and connectivities, and with a set of the
most commonly used metrics);

\item package \emph{itensor}---calculations by using (abstract) indices.
\end{itemize}

This system duplicates the functionality of the Macsyma package and
actually seeks no further development. At present, the capabilities of
these packages can hardly be considered as satisfactory.

\subsection{Reduce}

Reduce [\url{http://www.reduce-algebra.com/}] \cite{gerdt:1991:reduce}
is one of the oldest
(among currently existing systems) general-purpose computer algebra
systems. In 2009, its license was replaced from commercial to a
BSD-type. However, it should be noted that this was rather behindhand
because the community at that time had dealt with other free computer
algebra systems, and thus the system benefited little from the
transition to a free license.

The basic system consists of:
\begin{itemize}
\item the package \emph{atensor} mentioned above;
\item the package \emph{redten}
  [\url{http://www.scar.utoronto.ca/~harper/redten.html}] designed for
component calculations.
\end{itemize}

\subsection{Maple}

Maple
[\url{http://www.maplesoft.com/products/Maple/index.aspx}] is a
commercial general-purpose computer algebra system involving also the
tools for numerical computations.

\begin{itemize}
\item Maple has built-in tools for manipulating with tensor
  components, which are not inferior to similar tools in other
  computer algebra systems. Actually, there are two packages:
  \begin{itemize}
  \item package \emph{tensor}, which was originally designed for
    addressing problems of the general theory of relativity and
    component calculations;

  \item the powerful package \emph{DifferentialGeometry} (which has
    currently been involved in the main system) involves the
    subpackage \emph{Tensor} designed also for component
    calculations. Its great advantage is the possibility to use both
    the tensor analysis and the whole power of differential geometry
    (for example, the use of symmetries of groups and Lie algebras);
  \end{itemize}
\item \emph{GRTensor II} [\url{http://grtensor.phy.queensu.ca/}]
  (GPL-license) is one of the most powerful package of component
  tensor calculations.
\end{itemize}

\subsection{Mathematica}

Mathematica [\url{http://www.wolfram.com/mathematica/}] is a commercial
general-purpose computer algebra system developed by the Wolfram
Research company. The system involves a large number of computational
and interactive tools (for creating mathematical textbooks).
\begin{itemize}
\item \emph{MathTensor}
  [\url{http://smc.vnet.net/MathTensor.html}]~\cite{mathtensor} is a
  commercial package designed primarily for algebraic manipulations
  with tensors.

\item \emph{Cartan (Tensors in Physics)}
  [\url{http://www.adinfinitum.no/cartan/}] is a commercial package
  designed primarily for calculations in the theory of general
  relativity. Since the calculations are performed in specific
  metrics, the package operates with component indices.

\item \emph{Ricci} [\url{http://www.math.washington.edu/~lee/Ricci/}]
  is a package that generally supports algebraic manipulations and has
  also some elements of component operations. This package is in a
  state of stagnation.

\item The set of packages \emph{xAct} [\url{http://www.xact.es/}]
  (GPL-license) covers operations with both abstract and component
  indices.
\end{itemize}

In the next section, we consider in more detail two computer algebra
systems: \emph{Cadabra} and \emph{Maxima}.

The most important criterion for the choice of these two systems was
their license: only free computer algebra systems had been
considered. Thus, we excluded from consideration the extension packs
to commercial computer algebra systems (regardless of the license for
packages themselves).

Currently, the most advanced free computer algebra system for tensor
manipulations is \emph{Cadabra}. However, this system has not yet supported
component calculations. Therefore, we took the system \emph{Maxima} as a
companion to it.

A possible candidate for consideration could be \emph{Axiom}, but this
system is currently broken into several parts that are not quite
compatible.

\section{Tensor Operations in Cadabra}

To demonstrate the capabilities of Cadabra, we consider different
operations over tensors in this system. Since the current version of
Cadabra cannot operate with components, component operations will not
be considered.

\subsection{Nonindex Computing}

Let us define commuting rules and check that they are satisfied:
\begin{verbatim}
{A,B}::Commuting.
{C,D}::AntiCommuting.
\end{verbatim}
\begin{itemize}
\item The case of commuting tensors:
\begin{verbatim}
B A;
\end{verbatim}
\begin{equation*}
1 := B A;
\end{equation*}
\begin{verbatim}
@prodsort!(%);
\end{verbatim}
\begin{equation*}
1 := A B;
\end{equation*}
\item The case of anticommuting tensor:
\begin{verbatim}
D C;
\end{verbatim}
\begin{equation*}
2 := D C;
\end{equation*}
\begin{verbatim}
@prodsort!(%);
\end{verbatim}
\begin{equation*}
2 := -C D;
\end{equation*}
\end{itemize}

\subsection{Holonomic Coordinates}

Let us show that in the case of agreed metric and connectivity, the
covariant derivative of the metric tensor is vanishing (see~\eqref{eq:3}).

We define a set of indices, metric, and partial derivative:
\begin{verbatim}
{a,b,c,d,e,f,g,h,i,j,
  k,l,m,n,o,p,q,r,s,t,u#}::Indices.
g_{a b}::Metric.
\partial_{#}::PartialDerivative.
\end{verbatim}

We write the covariant derivative in the Christoffel symbols and the
Christoffel symbols through the metric tensor, which just specifies
the consistency of metric with connectivity:
\begin{verbatim}
\nabla := \partial_{c}{g_{a b}} - 
  g_{a d}\Gamma^{d}_{b c} - 
  g_{d b}\Gamma^{d}_{a c};
\end{verbatim}
\begin{equation*}
\nabla:= {\partial}_{c}{{g}_{a b}}   - {g}_{a d} {\Gamma}^{d} _{b c} - {g}_{d b} {\Gamma}^{d} _{a c};
\end{equation*}
\begin{verbatim}
Gamma:=\Gamma^{a}_{b c} -> (1/2) g^{a d} 
  ( \partial_{b}{g_{d c}} + 
  \partial_{c}{g_{b d}} - 
  \partial_{d}{g_{b c}} );
\end{verbatim}
\begin{equation*}
Gamma:= {\Gamma}^{a} _{b c} \rightarrow \frac{1}{2}  {g}^{a d}
({\partial}_{b}{{g}_{d c}}   + {\partial}_{c}{{g}_{b d}}   -
{\partial}_{d}{{g}_{b c}}  );
\end{equation*}

We substitute the expression of the Christoffel symbol through the
metric tensor in the expression for the covariant derivative:
\begin{verbatim}
@substitute!(\nabla)(@(Gamma));
\end{verbatim}
\begin{multline*}
  \nabla:= {\partial}_{c}{{g}_{a b}}   - \frac{1}{2}  {g}_{a d}
  {g}^{d e} ({\partial}_{b}{{g}_{e c}}   + {\partial}_{c}{{g}_{b
      e}}   - {\partial}_{e}{{g}_{b c}}  ) - 
  {} \\ -
  \frac{1}{2}  {g}_{d b}
  {g}^{d e} ({\partial}_{a}{{g}_{e c}}   + {\partial}_{c}{{g}_{a
      e}}   - {\partial}_{e}{{g}_{a c}}  );
\end{multline*}
and open the brackets:
\begin{verbatim}
@distribute!(%);
\end{verbatim}
\begin{multline*}
  \nabla:= {\partial}_{c}{{g}_{a b}}   - \frac{1}{2}  {g}_{a d} {g}^{d
    e} {\partial}_{b}{{g}_{e c}}   - 
  \frac{1}{2}  {g}_{a d} {g}^{d e} {\partial}_{c}{{g}_{b e}} + 
  {} \\ +
  \frac{1}{2}  {g}_{a d} {g}^{d e}  {\partial}_{e}{{g}_{b c}} - 
  \frac{1}{2}  {g}_{d b} {g}^{d e}  {\partial}_{a}{{g}_{e c}}  - 
  {} \\ -
  \frac{1}{2}  {g}_{d b} {g}^{d e}
  {\partial}_{c}{{g}_{a e}}   + \frac{1}{2}  {g}_{d b} {g}^{d e}
  {\partial}_{e}{{g}_{a c}}  ;
\end{multline*}
We raise and drop the indices unless all metric tensors are
eliminated, as indicated by the double exclamation mark:
\begin{verbatim}
@eliminate_metric!!(%);
\end{verbatim}
\begin{multline*}
  \nabla:= {\partial}_{c}{{g}_{a b}}   - \frac{1}{2} 
  {\partial}_{b}{{g}_{a c}}   - \frac{1}{2}  {\partial}_{c}{{g}_{b
      a}}   + \frac{1}{2}  {\partial}_{a}{{g}_{b c}}   -
  {} \\ -
  \frac{1}{2}  {\partial}_{a}{{g}_{b c}}   - \frac{1}{2} 
  {\partial}_{c}{{g}_{a b}}   + \frac{1}{2}  {\partial}_{b}{{g}_{a
      c}}  ;
\end{multline*}
Then, we bring the expression into the canonical form and collect the
terms. The result is zero as expected:
\begin{verbatim}
@canonicalise!(%);
\end{verbatim}
\begin{multline*}
  \nabla:= {\partial}_{c}{{g}_{a b}} - \frac{1}{2}
  {\partial}_{b}{{g}_{a c}} - \frac{1}{2} {\partial}_{c}{{g}_{a b}} +
  {} \\ +
  \frac{1}{2} {\partial}_{a}{{g}_{b c}} - \frac{1}{2}
  {\partial}_{a}{{g}_{b c}} - \frac{1}{2} {\partial}_{c}{{g}_{a b}} +
  \frac{1}{2} {\partial}_{b}{{g}_{a c}} ;
\end{multline*}
\begin{verbatim}
@collect_terms!(%);
\end{verbatim}
\begin{equation*}
\nabla:= 0;
\end{equation*}

\subsection{$\gamma$-Matrices}

Cadabra has advanced tools for operating with $\gamma$-matrices of any
dimension. For definiteness, we consider $\gamma$-matrices of Dirac
4-spinors.

To simplify the calculations, we define a set of actions that are
executed after each operation:
\begin{verbatim}
::PostDefaultRules( @@prodsort!(%), 
  @@eliminate_kr!(%),
  @@canonicalise!(%), 
  @@collect_terms!(%) ).
\end{verbatim}

We define the indices and their running values:
\begin{verbatim}
{a,b,c,d,e,f}::Indices(vector).
{a,b,c,d,e,f}::Integer(0..3).
\end{verbatim}
The space dimension will be used for finding the track of the
Kronecker delta.

The $\gamma$-matrices are specified using the metric (see~\eqref{eq:5}):
\begin{verbatim}
\gamma_{#}::GammaMatrix(metric=g).
g_{a b}::Metric.
g_{a}^{b}::KroneckerDelta.
\end{verbatim}

Now, we demonstrate some symmetry identities satisfied by
$\gamma$-matrices.
\begin{itemize}
\item Clifford-Dirac equation~\eqref{eq:5}:
\begin{verbatim}
\gamma_{a} \gamma_{b} + 
  \gamma_{b} \gamma_{a};
\end{verbatim}
\begin{equation*}
1 := {\gamma}_{a} {\gamma}_{b} + {\gamma}_{b} {\gamma}_{a};
\end{equation*}

The algorithm \verb|@join| converts the pairwise products of
$\gamma$-matrices into the sum of $\gamma$-matrices of higher valences
(see~\eqref{eq:10}). The additional argument \verb|expand| indicates that the rules
of antisymmetrization for $\gamma$-matrices are taken into account:
\begin{verbatim}
@join!(%){expand};
\end{verbatim}
\begin{equation*}
1 := 2\, {g}_{a b};
\end{equation*}

\item Convolution of two $\gamma$-matrices~\eqref{eq:gamma:sym:1}:
\begin{verbatim}
\gamma^{a} \gamma_{a};
\end{verbatim}
\begin{equation*}
2 := {\gamma}^{a} {\gamma}_{a};
\end{equation*}
\begin{verbatim}
@join!(%){expand};
\end{verbatim}
\begin{equation*}
2 := 4;
\end{equation*}

\item Identity~\eqref{eq:gamma:sym:2}:
\begin{verbatim}
\gamma^{a} \gamma^{b} 
  \gamma^{c} \gamma_{a};
\end{verbatim}
\begin{equation*}
3 := {\gamma}^{a} {\gamma}^{b} {\gamma}^{c} {\gamma}_{a};
\end{equation*}
\begin{verbatim}
@join!!(%){expand};
\end{verbatim}
\begin{equation*}
3 := ({\gamma}^{a b} + {g}^{a b}) ( - {\gamma}_{a}\,^{c} + {g}_{a}\,^{c});
\end{equation*}
\begin{verbatim}
@distribute!(%);
\end{verbatim}
\begin{equation*}
3 :=  - {\gamma}^{b a} {\gamma}^{c}\,_{a} - 2\, {\gamma}^{b c} + {g}^{b c};
\end{equation*}
\begin{verbatim}
@join!!(%){expand};
\end{verbatim}
\begin{equation*}
3 := 4\, {g}^{b c};
\end{equation*}

\item Identity~\eqref{eq:gamma:sym:3}:
\begin{verbatim}
\gamma_{a} \gamma_{b};
\end{verbatim}
\begin{equation*}
4 := {\gamma}_{a} {\gamma}_{b};
\end{equation*}
\begin{verbatim}
@join!!(%){expand};
\end{verbatim}
\begin{equation*}
4 := {\gamma}_{a b} + {g}_{a b};
\end{equation*}

\item Identity~\eqref{eq:gamma:sym:4}:
\begin{verbatim}
\gamma_{a} \gamma_{b} \gamma_{c};
\end{verbatim}
\begin{equation*}
5 := {\gamma}_{a} {\gamma}_{b} {\gamma}_{c};
\end{equation*}
\begin{verbatim}
@join!!(%){expand};
\end{verbatim}
\begin{equation*}
5 := ({\gamma}_{a b} + {g}_{a b}) {\gamma}_{c};
\end{equation*}
\begin{verbatim}
@distribute!(%);
\end{verbatim}
\begin{equation*}
5 := {\gamma}_{a b} {\gamma}_{c} + {\gamma}_{c} {g}_{a b};
\end{equation*}
\begin{verbatim}
@join!!(%){expand};
\end{verbatim}
\begin{equation*}
5 := {\gamma}_{a b c} + {\gamma}_{a} {g}_{b c} - {\gamma}_{b} {g}_{a c} + {\gamma}_{c} {g}_{a b};
\end{equation*}

\item A more complex identity:
\begin{equation}
  \label{eq:13}
  \gamma_{a b} \gamma_{b c} 
  \gamma_{d e} \gamma_{e a} =
  - 4\, {\gamma}_{c d} + 21\, {g}_{c d} \hat{I}
\end{equation}
In Cadabra, this identity has the form
\begin{verbatim}
\gamma_{a b} \gamma_{b c} 
  \gamma_{d e} \gamma_{e a};
\end{verbatim}
\begin{equation*}
6 := -{\gamma}_{a b} {\gamma}_{c a} {\gamma}_{d e} {\gamma}_{b e};
\end{equation*}
\begin{verbatim}
@join!!(%){expand};
\end{verbatim}
\begin{equation*}
6 := -(2\, {\gamma}_{b c} + 3\, {g}_{b c}) (2\, {\gamma}_{b d} - 3\, {g}_{b d});
\end{equation*}
\begin{verbatim}
@distribute!(%);
\end{verbatim}
\begin{equation*}
6 :=  - 4\, {\gamma}_{c b} {\gamma}_{d b} - 12\, {\gamma}_{c d} + 9\, {g}_{c d};
\end{equation*}
\begin{verbatim}
@join!(%){expand};
\end{verbatim}
\begin{equation*}
6 :=  - 4\, {\gamma}_{c d} + 21\, {g}_{c d};
\end{equation*}
\end{itemize}

\subsection{Monoterm Symmetries}

As an example of monoterm symmetry, we consider the symmetries of
Riemann tensor~\eqref{eq:riman:1}. To this end, we first specify the symmetry
properties using Young diagrams:
\begin{verbatim}
R_{a b c d}::TableauSymmetry(shape={2,2}, 
  indices={0,2,1,3}).
\end{verbatim}
In this example, the symmetry has the form {\scriptsize\young(ac,bd)}.

Then, we perform symmetric and antisymmetric permutation of the
indices. The algorithm \verb|@canonicalise| brings the operand into the
canonical form, accounting for monoterm symmetries.
\begin{verbatim}
R_{c d a b};
\end{verbatim}
\begin{equation*}
1 := {R}_{c d a b};
\end{equation*}
\begin{verbatim}
@canonicalise!(%);
\end{verbatim}
\begin{equation*}
1 := {R}_{a b c d};
\end{equation*}
\begin{verbatim}
R_{a b c d} + R_{b a c d};
\end{verbatim}
\begin{equation*}
2 := {R}_{a b c d} + {R}_{b a c d};
\end{equation*}
\begin{verbatim}
@canonicalise!(%);
\end{verbatim}
\begin{equation*}
2 := {R}_{a b c d} - {R}_{a b c d};
\end{equation*}
\begin{verbatim}
@collect_terms!(%);
\end{verbatim}
\begin{equation*}
2 := 0;
\end{equation*}

\subsection{Multiterm Symmetries}

We demonstrate the operation with multiterm symmetries by an example
of the Riemann tensor.

We introduce notations for indices and derivatives:
\begin{verbatim}
{a,b,c,d,e,f,g#}::Indices(vector).
\nabla{#}::Derivative.
\end{verbatim}

The symmetry can be specified by the Young diagram
{\scriptsize\young(ace,bd)}, as in the previous case:
\begin{verbatim}
\nabla_{e}{R_{a b c d}}::TableauSymmetry( 
  shape={3,2}, indices={1,3,0,2,4} ).
\end{verbatim}

However, it is more convenient the following notation:
\begin{verbatim}
R_{a b c d}::RiemannTensor.
\end{verbatim}

Similarly, we deal with the covariant derivative of the Riemann
tensor, which satisfies the differential Bianchi identity:
\begin{verbatim}
\nabla_{e}{R_{a b c d}}::SatisfiesBianchi.
\end{verbatim}

Let us check first Bianchi identity~\eqref{eq:riman:2}.
\begin{verbatim}
R_{a b c d} + R_{a c d b} + R_{a d b c};
\end{verbatim}
\begin{equation*}
1 := R_{a b c d} + R_{a c d b} + R_{a d b c};
\end{equation*}
\begin{verbatim}
@young_project_tensor!2(%){ModuloMonoterm}:
\end{verbatim}
\begin{verbatim}
@collect_terms!(%);
\end{verbatim}
\begin{equation*}
1 := 0;
\end{equation*}

Now, we demonstrate that second (differential) Bianchi
identity~\eqref{eq:riman:3} is satisfied:
\begin{verbatim}
\nabla_{e}{R_{a b c d}} + 
  \nabla_{c}{R_{a b d e}} + 
  \nabla_{d}{R_{a b e c}};
\end{verbatim}
\begin{equation*}
2 := {\nabla}_{e}{{R}_{a b c d}}\,  + {\nabla}_{c}{{R}_{a b d e}}\,  + {\nabla}_{d}{{R}_{a b e c}}\, ;
\end{equation*}
\begin{verbatim}
@young_project_tensor!2(%){ModuloMonoterm}:
\end{verbatim}
\begin{verbatim}
@collect_terms!(%);
\end{verbatim}
\begin{equation*}
2 := 0;
\end{equation*}

\section{An Example of Tensor Calculations in Maxima}

Because Cadabra currently does not support component computations, we
demonstrate them in the Maxima system. As an example, we consider the
Maxwell equations written in cylindrical coordinates in a holonomic
basis~\cite{pfur-2012-1}.

First, we load a small package written by us that contains definitions
for differential operators:

\noindent
\begin{minipage}[t]{8ex}{\bf
\begin{verbatim}
(%i1) 
\end{verbatim}}
\end{minipage}
\begin{minipage}[t]{\textwidth}{
\begin{verbatim}
load("diffop.mac")$
\end{verbatim}}
\end{minipage}

We define a cylindrical coordinate system:

\noindent
\begin{minipage}[t]{8ex}{\bf
\begin{verbatim}
(%i2) 
\end{verbatim}}
\end{minipage}
\begin{minipage}[t]{\textwidth}{
\begin{verbatim}
ct_coordsys(polar cylindrical)$
\end{verbatim}}
\end{minipage}

We consider the components of the metric tensor $g_{ij}$:

\noindent
\begin{minipage}[t]{8ex}{\bf
\begin{verbatim}
(%i3) 
\end{verbatim}}
\end{minipage}
\begin{minipage}[t]{\textwidth}{
\begin{verbatim}
lg;
\end{verbatim}}
\end{minipage}
\begin{math}\displaystyle
\parbox{8ex}{(\%o3) }
\begin{pmatrix}1 & 0 & 0\cr 0 & {r}^{2} & 0\cr 0 & 0 & 1\end{pmatrix}
\end{math}

We calculate and consider the components of the metric tensor $g^{ij}$: 

\noindent
\begin{minipage}[t]{8ex}{\bf
\begin{verbatim}
(%i4) 
\end{verbatim}}
\end{minipage}
\begin{minipage}[t]{\textwidth}{
\begin{verbatim}
cmetric()$
\end{verbatim}}
\end{minipage}

\noindent
\begin{minipage}[t]{8ex}{\bf
\begin{verbatim}
(%i5) 
\end{verbatim}}
\end{minipage}
\begin{minipage}[t]{\textwidth}{
\begin{verbatim}
ug;
\end{verbatim}}
\end{minipage}
\begin{math}\displaystyle
\parbox{8ex}{(\%o5) }
\begin{pmatrix}1 & 0 & 0\cr 0 & \frac{1}{{r}^{2}} & 0\cr 0 & 0 & 1\end{pmatrix}
\end{math}

We define the required vectors and determine their coordinate
dependences. Due to the limitations of Maxima, we denote $j^1$ by \verb|j1|
and $j_1$ by \verb|j_1|:

\noindent
\begin{minipage}[t]{8ex}{\bf
\begin{verbatim}
(%i6) 
\end{verbatim}}
\end{minipage}
\begin{minipage}[t]{\textwidth}{
\begin{verbatim}
j:[j1,j2,j3]$
depends(j,cons(t,ct_coords))$
\end{verbatim}}
\end{minipage}

\noindent
\begin{minipage}[t]{8ex}{\bf
\begin{verbatim}
(%i8) 
\end{verbatim}}
\end{minipage}
\begin{minipage}[t]{\textwidth}{
\begin{verbatim}
B:[B1,B2,B3]$ 
depends(B,cons(t,ct_coords))$
\end{verbatim}}
\end{minipage}

\noindent
\begin{minipage}[t]{8ex}{\bf
\begin{verbatim}
(%i10) 
\end{verbatim}}
\end{minipage}
\begin{minipage}[t]{\textwidth}{
\begin{verbatim}
D:[D1,D2,D3]$
depends(D,cons(t,ct_coords))$
\end{verbatim}}
\end{minipage}

\noindent
\begin{minipage}[t]{8ex}{\bf
\begin{verbatim}
(%i12) 
\end{verbatim}}
\end{minipage}
\begin{minipage}[t]{\textwidth}{
\begin{verbatim}
H:[H_1,H_2,H_3]$ 
depends(H,cons(t,ct_coords))$
\end{verbatim}}
\end{minipage}

\noindent
\begin{minipage}[t]{8ex}{\bf
\begin{verbatim}
(%i14) 
\end{verbatim}}
\end{minipage}
\begin{minipage}[t]{\textwidth}{
\begin{verbatim}
E:[E_1,E_2,E_3]$ 
depends(E,cons(t,ct_coords))$
\end{verbatim}}
\end{minipage}

Now, we calculate all sides of the Maxwell equations written in
cylindrical coordinates:

\noindent
\begin{minipage}[t]{8ex}{\bf
\begin{verbatim}
(%i16) 
\end{verbatim}}
\end{minipage}
\begin{minipage}[t]{\textwidth}{
\begin{verbatim}
Div(B);
\end{verbatim}}
\end{minipage}
\begin{math}\displaystyle
\parbox{8ex}{(\%o16) }
\frac{d}{d\,z}\,B3+\frac{d}{d\,\theta}\,B2+\frac{d}{d\,r}\,B1+\frac{B1}{r}
\end{math}

\noindent
\begin{minipage}[t]{8ex}{\bf
\begin{verbatim}
(%i17) 
\end{verbatim}}
\end{minipage}
\begin{minipage}[t]{\textwidth}{
\begin{verbatim}
Div(D) -4*%pi*rho;
\end{verbatim}}
\end{minipage}
\begin{math}\displaystyle
\parbox{8ex}{(\%o17) }
\frac{d}{d\,z}\,D3+\frac{d}{d\,\theta}\,D2+\frac{d}{d\,r}\,D1+\frac{D1}{r}-4\,\pi \,\rho
\end{math}

\noindent
\begin{minipage}[t]{8ex}{\bf
\begin{verbatim}
(%i18) 
\end{verbatim}}
\end{minipage}
\begin{minipage}[t]{\textwidth}{
\begin{verbatim}
Rot(H) 
  + diff(transpose(matrix(D)),t)/c 
  - 4*%pi/c*transpose(matrix(j));
\end{verbatim}}
\end{minipage}
\begin{math}\displaystyle
\parbox{8ex}{(\%o18) }
\begin{pmatrix}
\frac{\frac{d}{d\,\theta}\,H\_3-\frac{d}{d\,z}\,H\_2}{\left| r\right|
}+\frac{\frac{d}{d\,t}\,D1}{c}-\frac{4\,\pi \,j1}{c}\cr
-\frac{\frac{d}{d\,r}\,H\_3-\frac{d}{d\,z}\,H\_1}{\left| r\right|
}+\frac{\frac{d}{d\,t}\,D2}{c}-\frac{4\,\pi \,j2}{c}\cr
\frac{\frac{d}{d\,r}\,H\_2-\frac{d}{d\,\theta}\,H\_1}{\left| r\right|
}+\frac{\frac{d}{d\,t}\,D3}{c}-\frac{4\,\pi \,j3}{c}
\end{pmatrix}
\end{math}

\noindent
\begin{minipage}[t]{8ex}{\bf
\begin{verbatim}
(%i19) 
\end{verbatim}}
\end{minipage}
\begin{minipage}[t]{\textwidth}{
\begin{verbatim}
Rot(E) 
  + diff(transpose(matrix(B)),t)/c;
\end{verbatim}}
\end{minipage}
\begin{math}\displaystyle
\parbox{8ex}{(\%o19) }
\begin{pmatrix}
\frac{\frac{d}{d\,\theta}\,E\_3-\frac{d}{d\,z}\,E\_2}{\left| r\right|
}+\frac{\frac{d}{d\,t}\,B1}{c}\cr
\frac{\frac{d}{d\,t}\,B2}{c}-\frac{\frac{d}{d\,r}\,E\_3-\frac{d}{d\,z}\,E\_1}{\left|
    r\right| }\cr
\frac{\frac{d}{d\,r}\,E\_2-\frac{d}{d\,\theta}\,E\_1}{\left| r\right|
}+\frac{\frac{d}{d\,t}\,B3}{c}
\end{pmatrix}
\end{math}

Thus, the result coincided with the analytical expressions obtained
in~\cite{pfur-2012-1}.

\section{Conclusions}

This paper was prompted by the fact that the authors intended to
conduct bulky tensor calculations in computer algebra systems. The
search for an appropriate system brought them to formulate a set of
criteria that must be met by such a computer algebra system.

Because there are several types of tensor calculations, the full
implementation of tensor calculations in computer algebra systems
requires a broad set of capabilities. Unfortunately, at present, there
are almost no systems with a fully satisfactory support of tensors.

Component tensor calculations require almost no additional features of
the universal computer algebra system. Therefore, the packages
implementing this functionality are used most widely (for example,
\emph{Maxima}, \emph{Maple}, and \emph{Mathematica}).

The tensor notation is very different from the usual functional
notation that is used by the vast majority of computer algebra
systems. Therefore, the efficient operation with tensors would require
a specialized system (or a specialized add-on over the universal
system) that supports the natural tensor notation (for example,
\emph{Cadabra}).

As a result, we failed to find a system that fully meets the needs of
tensor calculus. At the moment, the authors use the \emph{Cadabra}
specialized system and a universal computer algebra system (currently,
\emph{Maxima}).

\bibliographystyle{abbrvnat}

\bibliography{bib/cas_tensor}


\end{document}


\title{Тензорные расчёты в системах компьютерной алгебры}

\author{А. В. Королькова}
\email{akorolkova@sci.pfu.edu.ru}
\author{Д. С. Кулябов}
\email{yamadharma@gmail.com}
\author{Л. А. Севастьянов}
\email{leonid.sevast@gmail.com}

\affiliation{Кафедра систем телекоммуникаций\\
  Российский университет дружбы народов\\
  ул. Миклухо-Маклая, д.6, Москва, 117198, Россия}

\thanks{Опубликовано в: 
  A.~V. Korol’kova, D.~S. Kulyabov, and L.~A. Sevast’yanov.
 {Tensor computations in computer algebra systems}.
 \emph{Programming and Computer Software}, 39\penalty0 (3):\penalty0
 135--142, 2013.
 ISSN 0361-7688.
 \href{http://dx.doi.org/10.1134/S0361768813030031}{doi:~10.1134/S0361768813030031}.}

\thanks{Исходные тексты:
  \url{https://bitbucket.org/yamadharma/articles-2011-cas_tensor}}

\begin{abstract}
  В статье рассмотрены три вида тензорных расчётов. В соответствии с
  ними авторы попытались сформулировать критерии, которым должна
  удовлетворять система компьютерной алгебры для работы с тензорами.
  Сделан краткий обзор текущего состояния тензорных вычислений в
  разных системах компьютерной алгебры. Тензорные расчёты
  проиллюстрированы соответствующими примерами, реализованными в
  конкретных системах: Cadabra и Maxima.
\end{abstract}

\maketitle

\section{Введение}




Тензорные вычисления используются во многих областях физики. Следует
заметить, что во всей своей мощи формализм тензорного анализа
проявляется не во всех областях, достаточно часто используют его
упрощённые варианты.

Каждая тензорная операция сама по себе достаточно проста. Однако даже
при стандартных вычислениях приходится выполнять множество
элементарных операций. Эти операции требуют большой внимательности и
скрупулёзности. Именно поэтому в данной области актуальны разные
упрощения нотации,  оптимизация операций (например, тензорные
диаграммы Пенроуза). 

Одной из задач систем компьютерной алгебры является освобождение
исследователя от рутинных операций, что актуально и в случае
тензорного исчисления.

\section{Основные области применения и типы записи тензоров}

Чтобы определить основные виды операции с тензорами, рассмотрим
основные области их применения.

\subsection{Безындексные вычисления для теоретических построений}

Безындексные вычисления обычно применяются в теоретических построениях
и часто противопоставляются компонентным вычислениям. Посмотрим, как
можно реализовать основные тензорные операции в безындексном случае.

\begin{itemize}

\item \emph{Сложение тензоров}. При сложении двух тензоров валентности 
$\left[\begin{smallmatrix} p\\q \end{smallmatrix}\right]$ получаем
тензор валентности $\left[\begin{smallmatrix}
    p\\q \end{smallmatrix}\right]$:
\begin{equation}
  \label{eq:9}
  A + B = C.
\end{equation}
Сложение тензоров задаёт структуру абелевой группы.

\item \emph{Тензорное умножение}. При тензорном умножении тензора $A$ с
  валентностью $\left[\begin{smallmatrix}
      p\\q \end{smallmatrix}\right]$ 
на тензор $D$ с валентностью
$\left[\begin{smallmatrix} r\\s \end{smallmatrix}\right]$
получаем тензор $E$ с валентностью
$\left[\begin{smallmatrix} p+r\\q+s \end{smallmatrix}\right]$:
\begin{equation}
  \label{eq:11}
  A \otimes D = E.
\end{equation}
Тензорное умножение задаёт структуру некоммутативной полугруппы.

\item \emph{Операция свёртывания}. Обозначим операцию свёртывания
  тензоров по
  последним индексам через $\mathfrak{C}$. Тогда под действием этой
  операции тензор $F$ с валентностью $\left[\begin{smallmatrix}
      p+1\\q+1 \end{smallmatrix}\right]$ переходит в тензор $G$ с
  валентностью $\left[\begin{smallmatrix}
      p\\q \end{smallmatrix}\right]$:
\begin{equation}
  \label{eq:12}
  \mathfrak{C} F = G.
\end{equation}

\item \emph{Операция перестановки индексов}. Данная операция
  необходима для задания симметрии тензоров (например, коммутатора или
  антикоммутатора тензора), для расширения операции свёртывания на
  свёртку по произвольным индексам. Однако в рамках безындексного
  подхода обозначить эту операцию
  нельзя. 
  Впрочем, простейшие симметрии мы можем явно указать в описании
  объекта (при этом для однозначности придётся наложить ограничения на
  валентность).

\end{itemize}

\subsection{Векторные вычисления}

Векторное исчисление --- простейший вариант тензорного исчисления
(вектор --- тензор валентности один). Вектор $a^{n}$ размерности $N$
представляется как совокупность набора компонент $n=\overline{1,N}$,
зависящего от базиса, и линейного закона преобразования компонент при
изменении базиса.  Часто используемые операции --- построение
разнообразных дифференциальных операторов и замена базиса. Наиболее
распространённые операторы: градиент, дивергенция, ротор (специфична
для трёхмерного пространства
$\mathbb{R}^3$)~\cite{pfur-2012-1,pfur-2011-2}).

Для компонентных расчётов необходимо определить базис, метрику и
связность (а соответственно и ковариантную производную). В векторном
исчислении широкое распространение получил голономный базис, который
строится как совокупность частных производных от координат в
касательном расслоении и дуального базиса как 1-формы в кокасательном
расслоении:
\begin{equation}
  \label{eq:2}
  \vec{\delta}_{i} = \frac{\partial}{\partial x^{i}},
  \quad
  \vec{\delta}^{i} = d x^{i}.
\end{equation}
Связность и метрика строятся таким образом, чтобы ковариантная
производная от метрики равнялась нулю:
\begin{equation}
  \label{eq:3}
  \nabla_{k} g_{i j} = 0.
\end{equation}
В этом случае связность и метрика согласованы~\cite{sardanashvili-2}.

Следует заметить, что в векторных вычислениях   также часто используется
специальный неголономный базис, позволяющий не различать
контравариантные и ковариантные вектора,  сохранять размерность
при замене координат\footnote{При преобразовании длина переходит в
  длину, угол в угол и т.д.} (подробнее см. в
работе~\cite{pfur-2012-1}):
\begin{equation}
  \label{eq:4}
  \vec{\delta}_{i} = \frac{\partial}{\partial s^{i}},
  \quad
  \vec{\delta}^{i} = d s^{i},
  \quad
  d s^{i} = h_{j}^{i} d x^{j}.
\end{equation}
Здесь $d s^{i}$ --- элемент длины по соответствующей координате,
$h_{j}^{i}$~--- коэффициенты неголономности (в случае ортогональных
координат~--- коэффициенты Ламе).

\subsection{Дираковские 4-спиноры}

Специальным случаем тензорных объектов являются спиноры (называемые
также спин-тензорами). В частности спиноры являются представлениями
группы Лоренца с полуцелым старшим весом. Обычные тензоры являются
представлениями с целочисленным старшим весом.

По историческим причинам наиболее часто в исследованиях используются
дираковские 4-спиноры, которые применяют для записи уравнений Дирака,
описывающих фермионы со спином~$\frac{1}{2}$.  Дираковские 4-спиноры
суть неприводимые спиноры для случая $n = 4$ и $s=\pm 2$, где $n$~---
размерность векторного пространства, $s = n - 2u$~--- его сигнатура,
$u$~--- число отрицательных значений диагонального метрического
тензора $g_{a b}$.

Обычно для манипуляции с дираковскими спинорами используют
$\gamma$-матрицы, получаемые из уравнения
Клиффорда--Дирака~\cite{cartan}:
\begin{equation}
  \label{eq:5}
  \gamma_{(a} \gamma_{b)} = g_{a b} \hat{I},
\end{equation}
где $\gamma_{a}$ --- матрицы $N \times N$, $g_{a b}$~--- метрический
тензор, $\hat{I}$~--- единичная матрица $N \times N$, $N$~---
размерность спинорного пространства:
\begin{equation}
  \label{eq:7}
  N = 
  \begin{cases}
    2^{n/2}, &\text{чётное $n$,} \\
    2^{n/2 - 1/2}, &\text{нечётное $n$.}
  \end{cases}
\end{equation}
$\gamma$-матрицы являются элементами алгебры Клиффорда, порождающими
линейное преобразование спинового пространства. 

Поскольку $\gamma$-матрицы можно рассматривать как коэффициенты
перехода от спинового пространства к векторному, то более строго
следует ввести спиновые коэффициенты и записать уравнение~\eqref{eq:5}
следующим образом: 
\begin{equation}
  \label{eq:6}
  \gamma_{a \rho}^{\sigma} \gamma_{b \sigma}^{\tau} + \gamma_{b
    \rho}^{\sigma} \gamma_{a \sigma}^{\tau} = 2 g_{a b} \delta_{\rho}^{\sigma}.
\end{equation}

Для построения полной алгебры Клиффорда необходимы ещё и произведения
$\gamma$-матриц, однако в силу~\eqref{eq:5} достаточно рассматривать
только антисимметризованные произведения:
\begin{equation}
  \label{eq:10}
  \gamma_{a b \dots d} := \gamma_{[a} \gamma_{b} \cdots \gamma_{d]}. 
\end{equation}
Также вводится элемент $\gamma_{5}$:
\begin{equation}
  \label{eq:8}
  \gamma_{5} := \frac{i}{4!} e^{a b c d} \gamma_{a}\gamma_{b}\gamma_{c}\gamma_{d},
\end{equation}
где $e^{a b c d}$~--- альтернирующий тензор.

Манипуляции с $\gamma$-матрицами сводятся к набору соотношений,
следующих из алгебраических симметрий, например
\begin{gather}
    \gamma^a \gamma_a = 4 \hat{I}, 
    \label{eq:gamma:sym:1} \\
    \gamma^a \gamma^b \gamma^c \gamma_a = 4 g^{b c} \hat{I} ,
    \label{eq:gamma:sym:2} \\
    \gamma_{a} \gamma_{b} = \gamma_{ab} + g_{ab} \hat{I},
    \label{eq:gamma:sym:3} \\
    \gamma_a \gamma_b \gamma_c = \gamma_{a b c} +
    g_{a b} \gamma_c + g_{b c}\gamma_a -
    g_{a c} \gamma_b.   
    \label{eq:gamma:sym:4}
\end{gather}

\subsection{Тензорные вычисления в общей теории относительности}

Общая теория относительности стала первой физической теорией,
потребовавшей всю мощь дифференциальной геометрии и тензорных
вычислений~\cite{gerdt:1980:ufn}. В вычислениях возникают громоздкие
тензорные конструкции, которые можно упрощать, учитывая симметрии
тензоров.  Обычно выделяют одноэлементные (monoterm) и многоэлементные
(multiterm) симметрии.  Одним из основных элементов теории является
тензор Римана, обладающий как простейшими одноэлементными, так и
сложными многоэлементными симметриями типа тождеств Бьянки.

Одноэлементные симметрии соответствуют простым перестановочным
симметриям и задаются группой перестановок. Для тензора Римана,
например,  имеем:
\begin{equation}
  \label{eq:riman:1}
  R_{bacd} = - R_{abcd}, \quad R_{cdab} = R_{abcd}.
\end{equation}

Многоэлементные симметрии задаются алгеброй перестановок.
Тождество Бьянки имеет вид\footnote{Круглые скобки в~\eqref{eq:riman:2}
  обозначают симметризацию.}:
\begin{equation}
  \label{eq:riman:2}
R_{a(bcd)} = R_{abcd} + R_{acdb} + R_{adbc} = 0.
\end{equation}
Дифференциальное (второе) тождество Бьянки имеет вид\footnote{Точка с
  запятой в~\eqref{eq:riman:3} означает ковариантную производную.}:
\begin{equation}
  \label{eq:riman:3}
 R_{ab(cd;e)} = \nabla_{e} R_{abcd} + \nabla_{c} R_{abde} + \nabla_{d} R_{abec} = 0.
\end{equation}

Симметрии наиболее естественно задавать с помощью диаграмм
Юнга~\cite{barut:1980}.
Причём наличие предопределённых классов тензоров не отменяет
необходимость в явном задании симметрии. Например, тензор Римана
$R_{abcd}$ в разных источниках имеет симметрии
{\scriptsize\young(ac,bd)} или {\scriptsize\young(ab,cd)}.

\subsection{Типы записи тензоров}

Таким образом, опираясь на рассмотренные выше виды тензорных
вычислений, можно выделить три типа записи тензоров: компонентная
запись, запись с абстрактными индексами и безындексная запись.  Каждый
тип имеет свою специфику и  область применения.

Компонентные индексы, фактически, превращают тензор в набор скалярных
величин, применяемых при конкретных расчётах. Обычно оперировать с
компонентными индексами есть смысл лишь после упрощения тензорного
выражения и учёта всех его симметрий.

Безындексную запись часто используют, если исследователя интересует не
конечный результат, а симметрии тензоров. Однако эта форма записи
страдает недостатком выразительности: тензор рассматривается как
целостный объект, соответственно и симметрии возможно рассматривать
лишь те, которые относятся к тензору в целом. Для работы с объектами
сложной структуры приходится изобретать новые обозначения либо
добавлять словесные пояснения. Эту проблему и должны снять абстрактные
индексы~\cite{penrose}.

Абстрактные индексы следует рассматривать как усовершенствование
безындексной записи тензора.  Абстрактный индекс обозначает лишь
принадлежность тензора к определённому пространству, а не следование
тензорному правилу преобразования (в отличие от компонентных
индексов). В этом случае возможно рассмотрение как симметрий,
охватывающих весь тензор (все его индексы), так и симметрий отдельных
групп индексов.

\section{Тензорные вычисления и системы компьютерной алгебры}

 Современные системы компьютерной алгебры способны решать задачи
 достаточно широкого спектра и из разных областей знаний. Есть системы
 как узко специализированные, так и с претензией на универсальность
 (обзоры некоторых систем см., например, в
 работах~\cite{cain,pfur-2007-1-2,ech}). Рассмотрим некоторые системы
 компьютерной алгебры, в которых в той или иной степени реализована
 возможность работы с тензорами. 

\subsection{Требования к системе компьютерной алгебры}

Три типа записи тензоров соответствуют трём видам тензорных
аналитических вычислений, что приводит к определённым требованиям,
предъявляемым системе компьютерной алгебры.

Безындексные вычисления манипулируют с тензорами как с целостными
алгебраическими объектами. В данном случае возможно либо задание
самого простейшего типа симметрии (объект является представлением
какой-либо группы или алгебры), либо использование объектов с заранее
заданной симметрией.

Абстрактные индексы требуют возможности задания сложных типов
симметрии, например, через диаграммы Юнга. 
Кроме того, необходимо уметь работать с немыми индексами, задавать
и учитывать их при приведении к каноническому виду.  Оба вида
абстрактных вычислений используют информацию о симметриях для
приведения к каноническому виду и упрощения тензорных выражений.

Компонентные индексы требуют по сути скалярной системы компьютерной
алгебры, возможно наличия простейших операций с матрицами. Фактически
задаётся конкретная система координат и метрика. Поскольку все
операции производятся компонентами, теряется информация о тензоре как
целостном объекте, об его симметриях. Поэтому все операции с
симметриями и приведение к каноническому виду должны быть выполнены на
предыдущем этапе исследования.

\subsection{Нотация}

Использование систем компьютерной алгебры зачастую предполагает
интерактивную работу пользователя. В этом случае удобство нотации
играет важнейшую роль. Следует заметить, что исторически
математическая запись тензоров следует нотации системы \TeX{}, а
именно тензор $T^{a}_{b}$ записывается как \verb|T^{a}_{b}|. Поэтому
использование такой нотации было бы вполне естественным. Такой подход
реализован в системе \emph{Cadabra}. Однако \emph{Cadabra}~---
специализированная система для тензорных вычислений.  При введении
тензорной нотации в системах компьютерной алгебры общего назначения
следует учитывать ограничения этих систем (например, знак \verb|^|
обычно зарезервирован и используется для возведения в степень).

Поскольку системы общего назначения работают с функциями, а основной
внутренней структурой данных является список, то  используется
функционально"=списочная нотация. В качестве имени функции может
задаваться имя тензора, а ковариантные и контравариантные индексы
задаются либо префиксом (например, как в \emph{xAct}):
\begin{verbatim}
T(a,-b),
\end{verbatim}
либо позиционно (например, как в \emph{Maxima}):
\begin{verbatim}
T([a],[b]).
\end{verbatim}
Возможно также использование ассоциативных списков, например таким
образом:
\begin{verbatim}
Tensor[Name["T"], Indices[Up[a], Down[b]]].
\end{verbatim}

Рассмотрим наиболее интересные для практического применения реализации
тензорных вычислений в разных системах компьютерной алгебры. 

\subsection{Cadabra}

Cadabra [\url{http://cadabra.phi-sci.com/}] относится к типу
специализированных систем компьютерной алгебры. Область её
специализации~--- теория поля. Поскольку сложные тензорные расчёты
являются неотъемлемой частью теории поля, неудивительно, что поддержка
тензорных расчётов находится в этой системе на высоком уровне.

Однако в теории поля оперируют в основном абстрактными индексами,
компонентным вычислениям уделяется гораздо меньше внимания. Именно
поэтому, наверное, компонентные вычисления пока ещё не реализованы в
Cadabra. Хотя данная возможность и стоит в плане на реализацию.

Впрочем, компонентные вычисления требуют от системы компьютерной
алгебры функционала системы общего назначения, который отсутствует в
системе Cadabra.

\subsection{Maxima}

Maxima [\url{http://maxima.sourceforge.net/}]~--- одна из основных
свободных систем компьютерной алгебры общего назначения. Maxima
произошла от системы Macsyma, разрабатывавшейся в MIT с 1968 по 1982
годы.

В Maxima реализованы все три типа тензорных вычислений~\cite{toth}:
\begin{itemize}
\item пакет \emph{atensor}~--- безындексные алгебраические вычисления
  (в него заложен набор основных алгебр; основное предназначение ---
  упрощение тензорных выражений при помощи манипуляций как с
  одноэлементными, так и с многоэлементными симметриями);
\item пакет \emph{ctensor}~--- компонентные вычисления (есть
  возможность манипулирования с метрикой, связностями; в пакет включён
  набор наиболее употребимых метрик);
\item пакет \emph{itensor}~--- вычисления с использованием
  (абстрактных) индексов.
\end{itemize}

Реализация дублирует функциональность пакета Macsyma и фактически не
развивается. Возможности данных пакетов вряд ли можно считать
удовлетворительными на сегодняшний день.

\subsection{Reduce}

Reduce [\url{http://www.reduce-algebra.com/}],
\cite{gerdt:1991:reduce}~--- одна из старейших (из ныне живых) систем
компьютерной алгебры общего назначения. В 2009 году сменила лицензию с
коммерческой на лицензию BSD-типа. Правда, заметим, сделано это было
достаточно поздно, сообщество в это время занималось другими
свободными системами компьютерной алгебры, и поэтому большого
преимущества от перехода к свободной лицензии система не получила.

В основную систему входят:
\begin{itemize}
\item пакет \emph{atensor}, упомянутый ранее;
\item пакет \emph{redten}
  [\url{http://www.scar.utoronto.ca/~harper/redten.html}] предназначен
  для компонентных вычислений.
\end{itemize}

\subsection{Maple}

Система Maple
[\url{http://www.maplesoft.com/products/Maple/index.aspx}] является
коммерческой системой компьютерной алгебры общего назначения. Имеет в
своём составе также и средства для численных вычислений.

\begin{itemize}
\item Система Maple имеет встроенные средства для манипуляции с
  тензорными компонентами, которые не уступают аналогичным средствам в
  других системах компьютерной алгебры. Фактически это два пакета:
  \begin{itemize}
  \item пакет \emph{tensor} изначально был направлен на решение задач
    общей теории относительности и предназначен для компонентных
    вычислений;
  \item в рамках мощного пакета \emph{DifferentialGeometry} (вошедшего
    на данный момент в состав основной системы) присутствует подпакет
    \emph{Tensor}, предназначенный также для компонентных
    вычислений. Его большим преимуществом является возможность
    использования не только тензорного анализа, но и всей мощи
    дифференциальной геометрии (например, использование симметрий
    групп и алгебр Ли);
  \end{itemize}

\item пакет \emph{GRTensor II} [\url{http://grtensor.phy.queensu.ca/}]
  (лицензия GPL) является одним из самых мощных пакетов компонентных
  тензорных вычислений.
\end{itemize}

\subsection{Mathematica}

Mathematica [\url{http://www.wolfram.com/mathematica/}] ---
коммерческая система компьютерной алгебры общего назначения компании
Wolfram Research. Содержит в своём составе целый комплекс расчётных и
интерактивных инструментов (для создания математических учебных
пособий).

\begin{itemize}
\item Пакет \emph{MathTensor}
  [\url{http://smc.vnet.net/MathTensor.html}], \cite{mathtensor}~---
  коммерческий пакет, предназначенный в первую очередь для
  алгебраических манипуляций с тензорами.

\item Пакет \emph{Cartan (Tensors in Physics)}
  [\url{http://www.adinfinitum.no/cartan/}] --- коммерческий пакет,
  предназначенный в первую очередь для вычислений в общей теории
  относительности. Поскольку выполняются вычисления в конкретных
  метриках, пакет оперирует компонентными индексами.

\item Пакет \emph{Ricci}
  [\url{http://www.math.washington.edu/~lee/Ricci/}] в основном
  поддерживает алгебраические манипуляции, также имеет и некоторые
  элементы компонентных операций. Находится в состоянии стагнации.

\item Набор пакетов \emph{xAct} [\url{http://www.xact.es/}] (лицензия
  GPL) охватывает операции как с абстрактными, так и с компонентными
  индексами.
\end{itemize}

В следующем разделе  рассмотрим подробнее две системы компьютерной алгебры:
\emph{Cadabra} и \emph{Maxima}.

Важнейшим критерием выбора этих двух систем стала лицензия~---
рассматривались только свободные системы компьютерной алгебры. Таким
образом были исключены из рассмотрения пакеты расширения к
коммерческим системам компьютерной алгебры (вне зависимости от 
лицензии на сами пакеты).

Наиболее развитая на данный момент свободная система компьютерной
алгебры для тензорных манипуляций --- \emph{Cadabra}. Однако пока она
не поддерживает компонентные вычисления. Поэтому в качестве компаньона
к ней  выбрана система \emph{Maxima}. 

Возможным претендентом на рассмотрение могла стать система
\emph{Axiom}, однако на данный момент она распалась на несколько
не вполне совместимых ответвлений.

\section{Операции над тензорами в системе Cadabra}

В качестве демонстрации возможностей Cadabra рассмотрим
выполнение разных операций над тензорами в этой системе. Напомним, что
в текущей версии Cadabra отсутствует возможность работы с
компонентами, поэтому компонентные операции рассмотрены не будут.

\subsection{Безындексные вычисления}

Зададим правила коммутации и проверим, что они выполняются:
\begin{verbatim}
{A,B}::Commuting.
{C,D}::AntiCommuting.
\end{verbatim}
\begin{itemize}
\item Случай коммутирующих тензоров:
\begin{verbatim}
B A;
\end{verbatim}
\begin{equation*}
1 := B A;
\end{equation*}
\begin{verbatim}
@prodsort!(%);
\end{verbatim}
\begin{equation*}
1 := A B;
\end{equation*}
\item Случай антикоммутирующих тензоров:
\begin{verbatim}
D C;
\end{verbatim}
\begin{equation*}
2 := D C;
\end{equation*}
\begin{verbatim}
@prodsort!(%);
\end{verbatim}
\begin{equation*}
2 := -C D;
\end{equation*}
\end{itemize}

\subsection{Голономные координаты}

Покажем, что в случае согласованных метрики и связности ковариантная производная от метрического тензора равна
нулю~(см.~\eqref{eq:3}).

Зададим набор индексов, метрику и частную производную:
\begin{verbatim}
{a,b,c,d,e,f,g,h,i,j,
  k,l,m,n,o,p,q,r,s,t,u#}::Indices.
g_{a b}::Metric.
\partial_{#}::PartialDerivative.
\end{verbatim}

Запишем ковариантную производную через символы Кристоффеля и
символы Кристоффеля через метрический тензор, чем собственно и зададим
согласование метрики со связностью:
\begin{verbatim}
\nabla := \partial_{c}{g_{a b}} - 
  g_{a d}\Gamma^{d}_{b c} - 
  g_{d b}\Gamma^{d}_{a c};
\end{verbatim}
\begin{equation*}
\nabla:= {\partial}_{c}{{g}_{a b}}   - {g}_{a d} {\Gamma}^{d} _{b c} - {g}_{d b} {\Gamma}^{d} _{a c};
\end{equation*}
\begin{verbatim}
Gamma:=\Gamma^{a}_{b c} -> (1/2) g^{a d} 
  ( \partial_{b}{g_{d c}} + 
  \partial_{c}{g_{b d}} - 
  \partial_{d}{g_{b c}} );
\end{verbatim}
\begin{equation*}
Gamma:= {\Gamma}^{a} _{b c} \rightarrow \frac{1}{2}  {g}^{a d}
({\partial}_{b}{{g}_{d c}}   + {\partial}_{c}{{g}_{b d}}   -
{\partial}_{d}{{g}_{b c}}  );
\end{equation*}

Подставим выражение символа Кристоффеля через метрический тензор в
выражение для ковариантной производной:
\begin{verbatim}
@substitute!(\nabla)(@(Gamma));
\end{verbatim}
\begin{multline*}
  \nabla:= {\partial}_{c}{{g}_{a b}}   - \frac{1}{2}  {g}_{a d}
  {g}^{d e} ({\partial}_{b}{{g}_{e c}}   + {\partial}_{c}{{g}_{b
      e}}   - {\partial}_{e}{{g}_{b c}}  ) - 
  {} \\ -
  \frac{1}{2}  {g}_{d b}
  {g}^{d e} ({\partial}_{a}{{g}_{e c}}   + {\partial}_{c}{{g}_{a
      e}}   - {\partial}_{e}{{g}_{a c}}  );
\end{multline*}
и раскроем скобки:
\begin{verbatim}
@distribute!(%);
\end{verbatim}
\begin{multline*}
  \nabla:= {\partial}_{c}{{g}_{a b}}   - \frac{1}{2}  {g}_{a d} {g}^{d
    e} {\partial}_{b}{{g}_{e c}}   - 
  \frac{1}{2}  {g}_{a d} {g}^{d e} {\partial}_{c}{{g}_{b e}} + 
  {} \\ +
  \frac{1}{2}  {g}_{a d} {g}^{d e}  {\partial}_{e}{{g}_{b c}} - 
  \frac{1}{2}  {g}_{d b} {g}^{d e}  {\partial}_{a}{{g}_{e c}}  - 
  {} \\ -
  \frac{1}{2}  {g}_{d b} {g}^{d e}
  {\partial}_{c}{{g}_{a e}}   + \frac{1}{2}  {g}_{d b} {g}^{d e}
  {\partial}_{e}{{g}_{a c}}  ;
\end{multline*}
Подымаем и опускаем индексы до тех пор, пока не уберём все метрические
тензоры, о чём говорит двойной восклицательный знак:
\begin{verbatim}
@eliminate_metric!!(%);
\end{verbatim}
\begin{multline*}
  \nabla:= {\partial}_{c}{{g}_{a b}}   - \frac{1}{2} 
  {\partial}_{b}{{g}_{a c}}   - \frac{1}{2}  {\partial}_{c}{{g}_{b
      a}}   + \frac{1}{2}  {\partial}_{a}{{g}_{b c}}   -
  {} \\ -
  \frac{1}{2}  {\partial}_{a}{{g}_{b c}}   - \frac{1}{2} 
  {\partial}_{c}{{g}_{a b}}   + \frac{1}{2}  {\partial}_{b}{{g}_{a
      c}}  ;
\end{multline*}
Далее приведём выражение к каноническому виду и приведём подобные. В
результате получим ноль, как и ожидалось:
\begin{verbatim}
@canonicalise!(%);
\end{verbatim}
\begin{multline*}
  \nabla:= {\partial}_{c}{{g}_{a b}} - \frac{1}{2}
  {\partial}_{b}{{g}_{a c}} - \frac{1}{2} {\partial}_{c}{{g}_{a b}} +
  {} \\ +
  \frac{1}{2} {\partial}_{a}{{g}_{b c}} - \frac{1}{2}
  {\partial}_{a}{{g}_{b c}} - \frac{1}{2} {\partial}_{c}{{g}_{a b}} +
  \frac{1}{2} {\partial}_{b}{{g}_{a c}} ;
\end{multline*}
\begin{verbatim}
@collect_terms!(%);
\end{verbatim}
\begin{equation*}
\nabla:= 0;
\end{equation*}

\subsection{$\gamma$-матрицы}

Cadabra имеет развитые средства для работы с $\gamma$-матрицами любой
размерности. Для определённости будем использовать $\gamma$-матрицы
дираковских 4-спиноров.

Для упрощения вычислений зададим набор действий, выполняющихся после
каждой операции:
\begin{verbatim}
::PostDefaultRules( @@prodsort!(%), 
  @@eliminate_kr!(%),
  @@canonicalise!(%), 
  @@collect_terms!(%) ).
\end{verbatim}

Зададим индексы и пробегаемые ими значения: 
\begin{verbatim}
{a,b,c,d,e,f}::Indices(vector).
{a,b,c,d,e,f}::Integer(0..3).
\end{verbatim}
Размерность пространства будет использоваться при нахождении следа
символа Кронекера.

При задании $\gamma$-матриц указывается метрика (см.~\eqref{eq:5}):
\begin{verbatim}
\gamma_{#}::GammaMatrix(metric=g).
g_{a b}::Metric.
g_{a}^{b}::KroneckerDelta.
\end{verbatim}

Теперь продемонстрируем несколько симметрийных тождеств, которым
удовлетворяют $\gamma$-матрицы.
\begin{itemize}
\item Уравнение Клиффорда--Дирака~\eqref{eq:5}:
\begin{verbatim}
\gamma_{a} \gamma_{b} + 
  \gamma_{b} \gamma_{a};
\end{verbatim}
\begin{equation*}
1 := {\gamma}_{a} {\gamma}_{b} + {\gamma}_{b} {\gamma}_{a};
\end{equation*}

Алгоритм \verb|@join| преобразует попарные произведения
$\gamma$-матриц в сумму $\gamma$-матриц высших валентностей
(см.~\eqref{eq:10}). Дополнительный аргумент \verb|expand| указывает
на то, что учитываются правила антисимметризации для $\gamma$-матриц:
\begin{verbatim}
@join!(%){expand};
\end{verbatim}
\begin{equation*}
1 := 2\, {g}_{a b};
\end{equation*}

\item Свёртка двух $\gamma$-матриц~\eqref{eq:gamma:sym:1}:
\begin{verbatim}
\gamma^{a} \gamma_{a};
\end{verbatim}
\begin{equation*}
2 := {\gamma}^{a} {\gamma}_{a};
\end{equation*}
\begin{verbatim}
@join!(%){expand};
\end{verbatim}
\begin{equation*}
2 := 4;
\end{equation*}

\item Тождество~\eqref{eq:gamma:sym:2}:
\begin{verbatim}
\gamma^{a} \gamma^{b} 
  \gamma^{c} \gamma_{a};
\end{verbatim}
\begin{equation*}
3 := {\gamma}^{a} {\gamma}^{b} {\gamma}^{c} {\gamma}_{a};
\end{equation*}
\begin{verbatim}
@join!!(%){expand};
\end{verbatim}
\begin{equation*}
3 := ({\gamma}^{a b} + {g}^{a b}) ( - {\gamma}_{a}\,^{c} + {g}_{a}\,^{c});
\end{equation*}
\begin{verbatim}
@distribute!(%);
\end{verbatim}
\begin{equation*}
3 :=  - {\gamma}^{b a} {\gamma}^{c}\,_{a} - 2\, {\gamma}^{b c} + {g}^{b c};
\end{equation*}
\begin{verbatim}
@join!!(%){expand};
\end{verbatim}
\begin{equation*}
3 := 4\, {g}^{b c};
\end{equation*}

\item Тождество~\eqref{eq:gamma:sym:3}:
\begin{verbatim}
\gamma_{a} \gamma_{b};
\end{verbatim}
\begin{equation*}
4 := {\gamma}_{a} {\gamma}_{b};
\end{equation*}
\begin{verbatim}
@join!!(%){expand};
\end{verbatim}
\begin{equation*}
4 := {\gamma}_{a b} + {g}_{a b};
\end{equation*}

\item Тождество~\eqref{eq:gamma:sym:4}:
\begin{verbatim}
\gamma_{a} \gamma_{b} \gamma_{c};
\end{verbatim}
\begin{equation*}
5 := {\gamma}_{a} {\gamma}_{b} {\gamma}_{c};
\end{equation*}
\begin{verbatim}
@join!!(%){expand};
\end{verbatim}
\begin{equation*}
5 := ({\gamma}_{a b} + {g}_{a b}) {\gamma}_{c};
\end{equation*}
\begin{verbatim}
@distribute!(%);
\end{verbatim}
\begin{equation*}
5 := {\gamma}_{a b} {\gamma}_{c} + {\gamma}_{c} {g}_{a b};
\end{equation*}
\begin{verbatim}
@join!!(%){expand};
\end{verbatim}
\begin{equation*}
5 := {\gamma}_{a b c} + {\gamma}_{a} {g}_{b c} - {\gamma}_{b} {g}_{a c} + {\gamma}_{c} {g}_{a b};
\end{equation*}

\item Более сложное тождество:
\begin{equation}
  \label{eq:13}
  \gamma_{a b} \gamma_{b c} 
  \gamma_{d e} \gamma_{e a} =
  - 4\, {\gamma}_{c d} + 21\, {g}_{c d} \hat{I}
\end{equation}
в Cadabra будет иметь вид:
\begin{verbatim}
\gamma_{a b} \gamma_{b c} 
  \gamma_{d e} \gamma_{e a};
\end{verbatim}
\begin{equation*}
6 := -{\gamma}_{a b} {\gamma}_{c a} {\gamma}_{d e} {\gamma}_{b e};
\end{equation*}
\begin{verbatim}
@join!!(%){expand};
\end{verbatim}
\begin{equation*}
6 := -(2\, {\gamma}_{b c} + 3\, {g}_{b c}) (2\, {\gamma}_{b d} - 3\, {g}_{b d});
\end{equation*}
\begin{verbatim}
@distribute!(%);
\end{verbatim}
\begin{equation*}
6 :=  - 4\, {\gamma}_{c b} {\gamma}_{d b} - 12\, {\gamma}_{c d} + 9\, {g}_{c d};
\end{equation*}
\begin{verbatim}
@join!(%){expand};
\end{verbatim}
\begin{equation*}
6 :=  - 4\, {\gamma}_{c d} + 21\, {g}_{c d};
\end{equation*}
\end{itemize}

\subsection{Одноэлементные симметрии}

В качестве примера одноэлементной симметрии рассмотрим симметрии
тензора Римана~\eqref{eq:riman:1}. Для этого зададим вначале
симметрийные свойства с помощью диаграммы Юнга:
\begin{verbatim}
R_{a b c d}::TableauSymmetry(shape={2,2}, 
  indices={0,2,1,3}).
\end{verbatim}
В~данном примере симметрия имеет вид {\scriptsize\young(ac,bd)}.

Далее произведём симметричную и антисимметричную перестановку
индексов. Алгоритм \verb|@canonicalise| приводит операнд к
канонической форме, учитывая одноэлементные симметрии.
\begin{verbatim}
R_{c d a b};
\end{verbatim}
\begin{equation*}
1 := {R}_{c d a b};
\end{equation*}
\begin{verbatim}
@canonicalise!(%);
\end{verbatim}
\begin{equation*}
1 := {R}_{a b c d};
\end{equation*}
\begin{verbatim}
R_{a b c d} + R_{b a c d};
\end{verbatim}
\begin{equation*}
2 := {R}_{a b c d} + {R}_{b a c d};
\end{equation*}
\begin{verbatim}
@canonicalise!(%);
\end{verbatim}
\begin{equation*}
2 := {R}_{a b c d} - {R}_{a b c d};
\end{equation*}
\begin{verbatim}
@collect_terms!(%);
\end{verbatim}
\begin{equation*}
2 := 0;
\end{equation*}

\subsection{Многоэлементные симметрии}

Работу с многоэлементными симметриями продемонстрируем на примере
тензора Римана. 

Введём обозначения для индексов и производной:
\begin{verbatim}
{a,b,c,d,e,f,g#}::Indices(vector).
\nabla{#}::Derivative.
\end{verbatim}

Симметрию можно задавать с помощью диаграммы Юнга {\scriptsize\young(ace,bd)}, как в предыдущем
случае:
\begin{verbatim}
\nabla_{e}{R_{a b c d}}::TableauSymmetry( 
  shape={3,2}, indices={1,3,0,2,4} ).
\end{verbatim}

Однако удобнее использовать следующую нотацию:
\begin{verbatim}
R_{a b c d}::RiemannTensor.
\end{verbatim}

Аналогично поступим и с ковариантной производной от тензора Римана,
удовлетворяющей дифференциальному тождеству Бьянки:
\begin{verbatim}
\nabla_{e}{R_{a b c d}}::SatisfiesBianchi.
\end{verbatim}

Проверим первое тождество Бьянки~\eqref{eq:riman:2}.
\begin{verbatim}
R_{a b c d} + R_{a c d b} + R_{a d b c};
\end{verbatim}
\begin{equation*}
1 := R_{a b c d} + R_{a c d b} + R_{a d b c};
\end{equation*}
\begin{verbatim}
@young_project_tensor!2(%){ModuloMonoterm}:
\end{verbatim}
\begin{verbatim}
@collect_terms!(%);
\end{verbatim}
\begin{equation*}
1 := 0;
\end{equation*}

Теперь продемонстрируем выполнение второго (дифференциального)
тождества Бьянки~\eqref{eq:riman:3}:
\begin{verbatim}
\nabla_{e}{R_{a b c d}} + 
  \nabla_{c}{R_{a b d e}} + 
  \nabla_{d}{R_{a b e c}};
\end{verbatim}
\begin{equation*}
2 := {\nabla}_{e}{{R}_{a b c d}}\,  + {\nabla}_{c}{{R}_{a b d e}}\,  + {\nabla}_{d}{{R}_{a b e c}}\, ;
\end{equation*}
\begin{verbatim}
@young_project_tensor!2(%){ModuloMonoterm}:
\end{verbatim}
\begin{verbatim}
@collect_terms!(%);
\end{verbatim}
\begin{equation*}
2 := 0;
\end{equation*}

\section{Пример тензорных вычислений в Maxima}

Поскольку Cadabra на данный момент не поддерживает компонентные
вычисления, продемонстрируем их
в системе Maxima. В качестве примера рассмотрим запись уравнений
Максвелла в цилиндрических координатах в голономном
базисе~\cite{pfur-2012-1}.

Загрузим вначале написанный нами небольшой пакет, содержащий
определения для дифференциальных операторов:

\noindent
\begin{minipage}[t]{8ex}{\bf
\begin{verbatim}
(%i1) 
\end{verbatim}}
\end{minipage}
\begin{minipage}[t]{\textwidth}{
\begin{verbatim}
load("diffop.mac")$
\end{verbatim}}
\end{minipage}

Зададим цилиндрическую систему координат:

\noindent
\begin{minipage}[t]{8ex}{\bf
\begin{verbatim}
(%i2) 
\end{verbatim}}
\end{minipage}
\begin{minipage}[t]{\textwidth}{
\begin{verbatim}
ct_coordsys(polar cylindrical)$
\end{verbatim}}
\end{minipage}

Посмотрим компоненты метрического тензора $g_{ij}$:

\noindent
\begin{minipage}[t]{8ex}{\bf
\begin{verbatim}
(%i3) 
\end{verbatim}}
\end{minipage}
\begin{minipage}[t]{\textwidth}{
\begin{verbatim}
lg;
\end{verbatim}}
\end{minipage}
\begin{math}\displaystyle
\parbox{8ex}{(\%o3) }
\begin{pmatrix}1 & 0 & 0\cr 0 & {r}^{2} & 0\cr 0 & 0 & 1\end{pmatrix}
\end{math}

Вычислим и посмотрим компоненты метрического тензора $g^{ij}$:

\noindent
\begin{minipage}[t]{8ex}{\bf
\begin{verbatim}
(%i4) 
\end{verbatim}}
\end{minipage}
\begin{minipage}[t]{\textwidth}{
\begin{verbatim}
cmetric()$
\end{verbatim}}
\end{minipage}

\noindent
\begin{minipage}[t]{8ex}{\bf
\begin{verbatim}
(%i5) 
\end{verbatim}}
\end{minipage}
\begin{minipage}[t]{\textwidth}{
\begin{verbatim}
ug;
\end{verbatim}}
\end{minipage}
\begin{math}\displaystyle
\parbox{8ex}{(\%o5) }
\begin{pmatrix}1 & 0 & 0\cr 0 & \frac{1}{{r}^{2}} & 0\cr 0 & 0 & 1\end{pmatrix}
\end{math}

Зададим необходимые векторы и определим их зависимость от
координат. Из-за ограничений Maxima обозначим $j^1$ через \verb|j1|, а
$j_1$ через \verb|j_1|:

\noindent
\begin{minipage}[t]{8ex}{\bf
\begin{verbatim}
(%i6) 
\end{verbatim}}
\end{minipage}
\begin{minipage}[t]{\textwidth}{
\begin{verbatim}
j:[j1,j2,j3]$
depends(j,cons(t,ct_coords))$
\end{verbatim}}
\end{minipage}

\noindent
\begin{minipage}[t]{8ex}{\bf
\begin{verbatim}
(%i8) 
\end{verbatim}}
\end{minipage}
\begin{minipage}[t]{\textwidth}{
\begin{verbatim}
B:[B1,B2,B3]$ 
depends(B,cons(t,ct_coords))$
\end{verbatim}}
\end{minipage}

\noindent
\begin{minipage}[t]{8ex}{\bf
\begin{verbatim}
(%i10) 
\end{verbatim}}
\end{minipage}
\begin{minipage}[t]{\textwidth}{
\begin{verbatim}
D:[D1,D2,D3]$
depends(D,cons(t,ct_coords))$
\end{verbatim}}
\end{minipage}

\noindent
\begin{minipage}[t]{8ex}{\bf
\begin{verbatim}
(%i12) 
\end{verbatim}}
\end{minipage}
\begin{minipage}[t]{\textwidth}{
\begin{verbatim}
H:[H_1,H_2,H_3]$ 
depends(H,cons(t,ct_coords))$
\end{verbatim}}
\end{minipage}

\noindent
\begin{minipage}[t]{8ex}{\bf
\begin{verbatim}
(%i14) 
\end{verbatim}}
\end{minipage}
\begin{minipage}[t]{\textwidth}{
\begin{verbatim}
E:[E_1,E_2,E_3]$ 
depends(E,cons(t,ct_coords))$
\end{verbatim}}
\end{minipage}

Теперь вычислим все части уравнений Максвелла в цилиндрических
координатах:

\noindent
\begin{minipage}[t]{8ex}{\bf
\begin{verbatim}
(%i16) 
\end{verbatim}}
\end{minipage}
\begin{minipage}[t]{\textwidth}{
\begin{verbatim}
Div(B);
\end{verbatim}}
\end{minipage}
\begin{math}\displaystyle
\parbox{8ex}{(\%o16) }
\frac{d}{d\,z}\,B3+\frac{d}{d\,\theta}\,B2+\frac{d}{d\,r}\,B1+\frac{B1}{r}
\end{math}

\noindent
\begin{minipage}[t]{8ex}{\bf
\begin{verbatim}
(%i17) 
\end{verbatim}}
\end{minipage}
\begin{minipage}[t]{\textwidth}{
\begin{verbatim}
Div(D) -4*%pi*rho;
\end{verbatim}}
\end{minipage}
\begin{math}\displaystyle
\parbox{8ex}{(\%o17) }
\frac{d}{d\,z}\,D3+\frac{d}{d\,\theta}\,D2+\frac{d}{d\,r}\,D1+\frac{D1}{r}-4\,\pi \,\rho
\end{math}

\noindent
\begin{minipage}[t]{8ex}{\bf
\begin{verbatim}
(%i18) 
\end{verbatim}}
\end{minipage}
\begin{minipage}[t]{\textwidth}{
\begin{verbatim}
Rot(H) 
  + diff(transpose(matrix(D)),t)/c 
  - 4*%pi/c*transpose(matrix(j));
\end{verbatim}}
\end{minipage}
\begin{math}\displaystyle
\parbox{8ex}{(\%o18) }
\begin{pmatrix}
\frac{\frac{d}{d\,\theta}\,H\_3-\frac{d}{d\,z}\,H\_2}{\left| r\right|
}+\frac{\frac{d}{d\,t}\,D1}{c}-\frac{4\,\pi \,j1}{c}\cr
-\frac{\frac{d}{d\,r}\,H\_3-\frac{d}{d\,z}\,H\_1}{\left| r\right|
}+\frac{\frac{d}{d\,t}\,D2}{c}-\frac{4\,\pi \,j2}{c}\cr
\frac{\frac{d}{d\,r}\,H\_2-\frac{d}{d\,\theta}\,H\_1}{\left| r\right|
}+\frac{\frac{d}{d\,t}\,D3}{c}-\frac{4\,\pi \,j3}{c}
\end{pmatrix}
\end{math}

\noindent
\begin{minipage}[t]{8ex}{\bf
\begin{verbatim}
(%i19) 
\end{verbatim}}
\end{minipage}
\begin{minipage}[t]{\textwidth}{
\begin{verbatim}
Rot(E) 
  + diff(transpose(matrix(B)),t)/c;
\end{verbatim}}
\end{minipage}
\begin{math}\displaystyle
\parbox{8ex}{(\%o19) }
\begin{pmatrix}
\frac{\frac{d}{d\,\theta}\,E\_3-\frac{d}{d\,z}\,E\_2}{\left| r\right|
}+\frac{\frac{d}{d\,t}\,B1}{c}\cr
\frac{\frac{d}{d\,t}\,B2}{c}-\frac{\frac{d}{d\,r}\,E\_3-\frac{d}{d\,z}\,E\_1}{\left|
    r\right| }\cr
\frac{\frac{d}{d\,r}\,E\_2-\frac{d}{d\,\theta}\,E\_1}{\left| r\right|
}+\frac{\frac{d}{d\,t}\,B3}{c}
\end{pmatrix}
\end{math}

Таким образом, результат  совпал с полученными в~\cite{pfur-2012-1}
аналитическими выражениями.

\section{Заключение}

Статья вызвана к жизни тем, что авторы хотели провести громоздкие
тензорные вычисления в системе компьютерной алгебры. Поиск
соответствующей системы заставил сформулировать набор критериев,
которым должна удовлетворять подобная система компьютерной алгебры.

Из-за наличия нескольких типов тензорных расчётов полная
реализация тензорных вычислений в системах компьютерной алгебры
требует широкого набора возможностей. К сожалению,
на данный момент
фактически отсутствуют системы с полностью удовлетворительной
поддержкой тензоров.

Компонентные тензорные вычисления практически не требуют
дополнительных возможностей от универсальной системы компьютерной
алгебры. Поэтому пакеты, реализующие данный функционал, представлены
наиболее широко (например, \emph{Maxima}, \emph{Maple},
\emph{Mathematica}).

Тензорная нотация сильно отличается от привычной функциональной
нотации, которую использует подавляющее большинство систем компьютерной
алгебры. Поэтому для эффективной работы с тензорами представляется
необходимым иметь специализированную систему (либо специализированную
надстройку над универсальной системой), поддерживающую естественную
тензорную нотацию (например, \emph{Cadabra}).

В результате не удалось найти систему, полностью удовлетворяющую
потребностям тензорного исчисления. Авторы пока остановились на наборе
из специализированной системы \emph{Cadabra} и универсальной системы
компьютерной алгебры (на данный момент \emph{Maxima}).

\bibliographystyle{gost2008l}

\bibliography{bib/cas_tensor}

